\documentclass[10pt,aps,prl,twocolumn,superscriptaddress,floatfix,longbibliography]{revtex4-1}

\usepackage{graphicx}
\usepackage{amsfonts}
\usepackage{amssymb}
\usepackage{amsmath}
\usepackage{txfonts}
\usepackage{lipsum}
\usepackage{color}
\usepackage{wasysym}
\usepackage[colorlinks=true, allcolors=blue]{hyperref}
\usepackage{bbold}
\usepackage{etoolbox} 
\usepackage[normalem]{ulem}
\usepackage{mathtools} 
\usepackage{multirow} 
\usepackage{hhline}
\usepackage{mathrsfs} 
\usepackage{soul}

\usepackage{letltxmacro}
\LetLtxMacro{\oldsqrt}{\sqrt}
\renewcommand{\sqrt}[2][\mkern8mu]{\mkern-6mu\mathop{}\oldsqrt[#1]{#2}}

\begin{document}

\title{
Extended regime of meta-stable metallic and insulating phases in a two-orbital electronic system
}

\author{M. Vandelli}
\affiliation{I. Institute of Theoretical Physics, University of Hamburg, Jungiusstrasse 9, 20355 Hamburg, Germany}
\affiliation{The Hamburg Centre for Ultrafast Imaging, Luruper Chaussee 149, 22761 Hamburg, Germany}
\affiliation{Max Planck Institute for the Structure and Dynamics of Matter,
Center for Free Electron Laser Science, 22761 Hamburg, Germany}

\author{J. Kaufmann}
\affiliation{Institute of Solid State Physics, TU Wien, 1040 Vienna, Austria}

\author{V. Harkov}
\affiliation{I. Institute of Theoretical Physics, University of Hamburg, Jungiusstrasse 9, 20355 Hamburg, Germany}
\affiliation{European X-Ray Free-Electron Laser Facility, Holzkoppel 4, 22869 Schenefeld, Germany}

\author{A. I. Lichtenstein}
\affiliation{I. Institute of Theoretical Physics, University of Hamburg, Jungiusstrasse 9, 20355 Hamburg, Germany}
\affiliation{European X-Ray Free-Electron Laser Facility, Holzkoppel 4, 22869 Schenefeld, Germany}
\affiliation{The Hamburg Centre for Ultrafast Imaging, Luruper Chaussee 149, 22761 Hamburg, Germany}

\author{K. Held}
\affiliation{Institute of Solid State Physics, TU Wien, 1040 Vienna, Austria}

\author{E. A. Stepanov}
\email{evgeny.stepanov@polytechnique.edu}
\affiliation{CPHT, CNRS, Ecole Polytechnique, Institut Polytechnique de Paris, F-91128 Palaiseau, France}

\begin{abstract}
We investigate the metal-to-insulator phase transition driven by the density-density electronic interaction in the quarter-filled model on a cubic lattice with two orbitals split by a crystal field.
We show that a systematic consideration of the non-local collective electronic fluctuations strongly affects the picture of the phase transition provided by the dynamical mean field theory. 
Our calculations reveal the appearance of metallic and Mott insulating states characterised by the same density but different values of the chemical potential, which is missing in the local approximation to electronic correlations. 
We find that the region of concomitant metastability of these two solutions is remarkably broad in terms of the interaction strength.
It starts at a critical value of the interaction slightly larger than the bandwidth and extends to more than twice the bandwidth, where the two solutions merge into a Mott insulating phase. 
Our results illustrate that non-local correlations can have crucial consequences on the electronic properties in the strongly correlated regime of the simplest multi-orbital systems.
\end{abstract}

\maketitle

There are two main mechanisms responsible for the formation of an insulating phase in electronic materials: a gap at the Fermi energy in the non-interacting band structure 
and the many-body localization induced by strong electronic interactions, as for instance the Mott scenario~\cite{Mott, RevModPhys.70.1039}.
The interplay between these different mechanisms can strongly affect the degree of electronic correlations and therefore the phase diagram of the material~\cite{PhysRevLett.78.507}.
Both these effects are especially important when a subset of doubly- or triply-degenerate localized orbitals appears in the electronic spectrum at Fermi energy.
Usually, the charge distribution on neighboring atoms lifts this degeneracy, which results in
a local splitting of the orbitals called crystal field splitting.
Strong electronic correlations may greatly renormalize the electronic spectral distribution, thus affecting the orbital splitting~\cite{10.1143/PTPS.160.233, PhysRevB.76.085127, PhysRevB.78.045115, PhysRevB.88.195116}.
The crystal field splitting also has a strong influence on the Mott transition in several materials, as it favors orbital polarization and orbital selective phenomena~\cite{Anisimov2002, PhysRevB.70.205116, PhysRevB.72.201102, de2009genesis, PhysRevLett.99.126405, Hackl_2009, PhysRevB.79.115119, PhysRevLett.102.126401,  PhysRevB.84.195130, PhysRevB.86.035150, Wang_2016, PhysRevB.93.155161, PhysRevB.94.075107, PhysRevB.100.115159}. 

The dynamical mean field theory (DMFT)~\cite{RevModPhys.68.13} is currently the most-widely used theoretical method for describing the Mott transition in realistic materials~\cite{RevModPhys.78.865, doi:10.1146/annurev-conmatphys-020911-125045}.
For instance, this method captures the coexistence of metallic and insulating phases that accompanies the Mott transition in both, single-band~\cite{PhysRevLett.70.1666, PhysRevB.48.7167, PhysRevB.49.10181, PhysRevLett.83.136, PhysRevB.75.125103, PhysRevLett.101.186403, PhysRevB.83.205136} and multi-orbital~\cite{PhysRevB.55.R4855, PhysRevB.66.165111, PhysRevLett.89.046401, PhysRevB.67.035119, PhysRevB.100.085104} systems.
However, in some cases DMFT is insufficient, because this theory accounts only for local correlation effects. 
Considering even short-range correlations beyond DMFT significantly modifies the coexistence region and drastically reduces the critical value of the interaction~\cite{PhysRevLett.101.186403}.
Long-range correlations can have even more dramatic consequences~\cite{PhysRevB.91.125109}.
Therefore, an important leap towards an accurate theoretical description of correlated materials would be to understand the effect of non-local collective electronic fluctuations on the spectral function.
Unfortunately, most of the available theoretical methods for multi-orbital systems are either limited to a weakly correlated regime~\cite{Bickers04, PhysRevB.69.104504, PhysRevB.75.224509, GW1,  GW2, GW3, RevModPhys.74.601},
or do not take into account all desired physical ingredients, such as long-range correlations~\cite{PhysRevB.84.020401, PhysRevB.89.195146, PhysRevB.91.235107, PhysRevLett.128.206401} or spatial magnetic fluctuations~\cite{PhysRevLett.90.086402, Tomczak_2012, PhysRevLett.109.237010, PhysRevB.88.165119, PhysRevB.88.235110, PhysRevB.90.165138, PhysRevLett.113.266403, PhysRevB.95.041112, PhysRevX.8.021038, Ryee2020}.
Attempts to go beyond these assumptions using diagrammatic methods lead to expensive numerical calculations~\cite{PhysRevLett.107.137007, PhysRevB.85.115128, PhysRevB.95.115107, doi:10.7566/JPSJ.87.041004, Boehnke_2018, acharya2019evening, PhysRevB.100.125120, PhysRevB.103.035120}, while unbiased quantum Monte Carlo methods are so far limited to specific parameter regimes or symmetries due to the fermionic sign problem~\cite{PhysRevLett.102.226402, PhysRevB.88.125108, PhysRevLett.110.107002, PhysRevB.99.235142, PhysRevLett.125.247001, PhysRevB.105.165124}.

In this work, we investigate the effect of non-local correlations on the Mott transition in a two-orbital model with the crystal field splitting and the density-density approximation for the  interaction.
This model is relevant for investigating the low-energy physics of some transition metal oxides~\cite{PhysRevB.70.205116} and of fulleride molecular crystals~\cite{PhysRevB.66.115107, Capone02, RevModPhys.81.943, PhysRevB.85.155452, PhysRevLett.118.177002}.
More importantly, this model is one of the simplest multi-orbital systems that allows for studying the influence of the orbital splitting on the Mott transition. 
So far this simple model has not been studied beyond the local DMFT approximation~\cite{PhysRevB.78.045115} due to computational difficulties associated with incorporating non-local correlations in the multi-orbital framework.
We challenge this solution of the problem by utilizing a relatively inexpensive diagrammatic extension of DMFT~\footnote{
For a review on diagrammatic extensions of DMFT see~\cite{RevModPhys.90.025003}.
} -- the dual triply irreducible local expansion (\mbox{D-TRILEX}) method~\cite{PhysRevB.100.205115, PhysRevB.103.245123, PhysRevLett.127.207205, arxiv.2204.06426}.
This approach accounts for the effect of the non-local collective electronic fluctuations on the spectral function in a self-consistent manner~\cite{PhysRevB.103.245123, stepanov2021coexisting, 2022arXiv220402895S}.
We find that, despite the apparent simplicity, the considered model displays a non-trivial behavior around the Mott transition.
In particular, considering the non-local correlations beyond DMFT reveals a broad coexistence region of meta-stable metallic and Mott insulating phases that extends from approximately the bandwidth to more than twice the bandwidth in the value of the interaction.
Our results might guide the understanding of the memristive effects experimentally observed in VO$_{2}$ thin-film samples \cite{doi:10.1063/1.1653835, doi:10.1063/1.3187531}.

{\it Method.}
The Hamiltonian of the considered two-orbital model on a cubic lattice
\begin{align*}
H =
\sum_{jj',l,\sigma} c^{\dagger}_{\hspace{-0.05cm}jl\sigma}\left(t^{l}_{\hspace{-0.05cm}jj'} + \Delta^{\phantom{l}}_{l}\,\delta^{\phantom{l}}_{\hspace{-0.05cm}jj'}\right) c^{\phantom{\dagger}}_{\hspace{-0.05cm}j'l\sigma}
+ \frac{U}{2}\hspace{-0.05cm}\sum_{j,ll'} n_{\hspace{-0.05cm}jl}\hspace{0.05cm} n_{\hspace{-0.05cm}jl'}
\end{align*}
contains three contributions.
We restrict the hopping to the nearest-neighbor lattice sites and set it to ${t^{l}_{\langle jj'\rangle}=1/6}$ for each of the two orbitals ${l\in\{1,2\}}$.
Hereinafter, the energy is expressed in units of the half-bandwidth of the cubic dispersion ${W/2=6t=1}$.
The interaction $U$ between electronic densities ${n_{\hspace{-0.05cm}jl} = \sum_{\sigma}c^{\dagger}_{\hspace{-0.05cm}jl\sigma}c^{\phantom{\dagger}}_{\hspace{-0.05cm}jl\sigma}}$ describes both the intra- and interorbital Coulomb repulsion.
Calculations are performed at quarter-filling, which corresponds to the average density of ${\langle n \rangle = 1}$ electron per two orbitals.
In order to induce an orbital polarization ${\delta n = (\langle n_{2}\rangle - \langle n_{1}\rangle})/\langle n \rangle$, we take a relatively large value for the crystal field splitting 
${\Delta = 2\Delta_1=-2\Delta_2} = 0.3$. 
This case was studied in details in Ref.~\cite{PhysRevB.78.045115} using DMFT. 
It was demonstrated, that local electronic correlations enlarge the orbital splitting, resulting in a high degree of orbital polarization.
Consequently, the single electron mostly populates the lower orbital (${l=2}$) that undergoes the Mott transition at a critical value of the electronic interaction. 
A similar interplay between the orbital polarization and Mott physics is also found in actual materials such as V$_2$O$_3$~\cite{PhysRevB.70.205116} and SrVO$_3$~\cite{Zhong2015, PhysRevB.94.201106, PhysRevMaterials.1.043803}, where it is important for the Mott transition. 

In order to investigate how non-local correlations affect the DMFT scenario of the Mott transition, we employ the \mbox{D-TRILEX} method~\cite{PhysRevB.100.205115, PhysRevB.103.245123, arxiv.2204.06426}, where collective electronic fluctuations are treated diagrammatically beyond DMFT.
This method was derived as an approximation to the dual boson theory~\cite{Rubtsov20121320, PhysRevB.90.235135, PhysRevB.93.045107, PhysRevB.94.205110, PhysRevLett.121.037204, Stepanov18-2, PhysRevB.99.115124, PhysRevB.100.165128, PhysRevB.102.195109, stepanov2021spin}, one of the most commonly used diagrammatic extensions of DMFT, cf.~\cite{RevModPhys.90.025003, PhysRevB.75.045118, PhysRevB.77.033101, PhysRevB.79.045133, PhysRevLett.102.206401,  PhysRevB.80.075104, PhysRevB.88.115112, PhysRevB.92.115109, PhysRevB.93.235124, PhysRevB.95.115107, PhysRevLett.119.166401, PhysRevB.102.195131, PhysRevB.103.035120, BRENER2020168310, PhysRevResearch.3.013149, PhysRevX.11.011058}.
The \mbox{D-TRILEX} method stands out for its lowered complexity, which allows one to address multi-band problems~\cite{PhysRevLett.127.207205, arxiv.2204.06426, 2022arXiv220402895S}, cf.~\cite{PhysRevB.95.115107, doi:10.7566/JPSJ.87.041004}, and its capability of correctly reproducing the results of more elaborate theories.
The reduction of the critical interaction for the Mott transition compared to DMFT~\cite{PhysRevB.100.205115} is very similar to cluster DMFT~\cite{PhysRevLett.101.186403}.
Additionally, it shows a precise agreement with exact benchmarks for some single- and multi-band systems~\cite{PhysRevB.103.245123, arxiv.2204.06426}.

If the system exhibits strong magnetic fluctuations, as frequently happens at half-filling, the Mott transition usually lies inside the antiferromagnetic (AFM) phase.
In this case, addressing the Mott transition requires to perform calculations in a symmetry broken phase, which is problematic.
Going away from half-filling suppresses the magnetic fluctuations and allows one to access the Mott transition from the paramagnetic phase.
According to our calculations, the highest critical temperature for the N\'eel transition for the considered quarter-filled model 
lies below ${T=0.06}$.
For this reason, we set the inverse temperature to ${T^{-1}=15}$, which ensures that the system is located outside the AFM phase but close to its boundary to observe strong magnetic fluctuations.
We perform DMFT calculations using the w2dynamics package~\cite{WALLERBERGER2019388}.
The \mbox{D-TRILEX} solution is based on the numerical implementation described in Ref.~\cite{arxiv.2204.06426}.
The local density of states (DOS) is obtained from the corresponding local Green's functions via analytical continuation using the ana\_cont package~\cite{kaufmann2021anacont}.

\begin{figure}[t!]
\includegraphics[width=1.\linewidth]{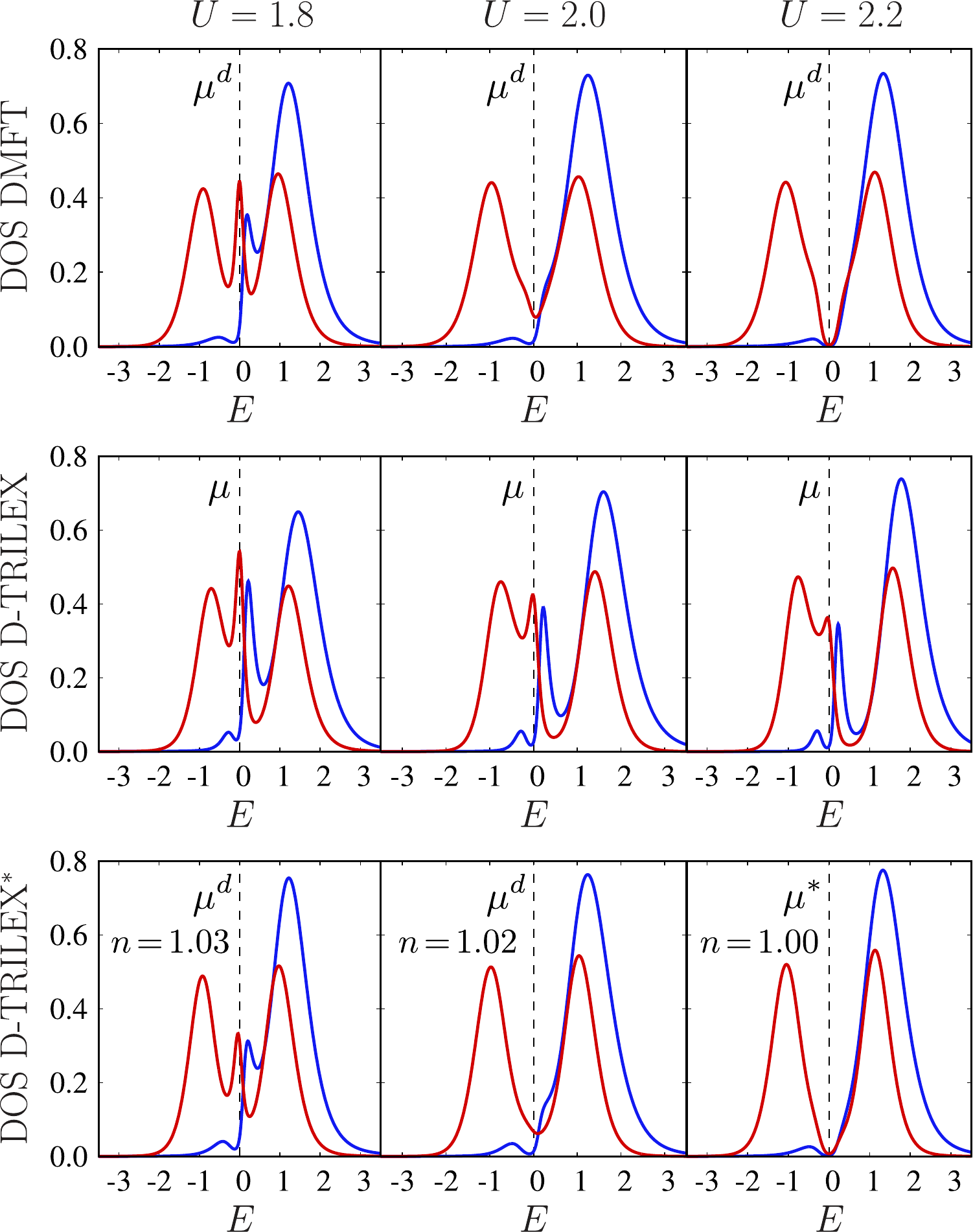}
\caption{DOS for the upper (${l=1}$, blue line) and lower (${l=2}$, red line) orbitals calculated for different interactions ${U=1.8}$ (left column), ${U=2.0}$ (middle column), and ${U=2.2}$ (right column).
Top row: DMFT solution at quarter-filling that corresponds to the chemical potential $\mu^{d}$.
Middle row: quarter-filled metallic \mbox{D-TRILEX} solution for the chemical potential $\mu$.
Bottom row: a further \mbox{D-TRILEX}$^{*}$ calculation based on the DMFT solution. 
Calculations for ${U=1.8}$ and ${U=2.0}$ are performed for $\mu^{d}$. The resulting ${\langle n \rangle > 1}$ is specified in panels.
At ${U=2.2}$ the quarter-filled D-TRILEX$^{*}$ solution appears at ${\mu^{*}\simeq\mu^{d}}$ and corresponds to the Mott insulating state.
\label{fig:DOS}}
\end{figure}

\begin{figure}[t!]
\centering
\includegraphics[width=1.0\linewidth]{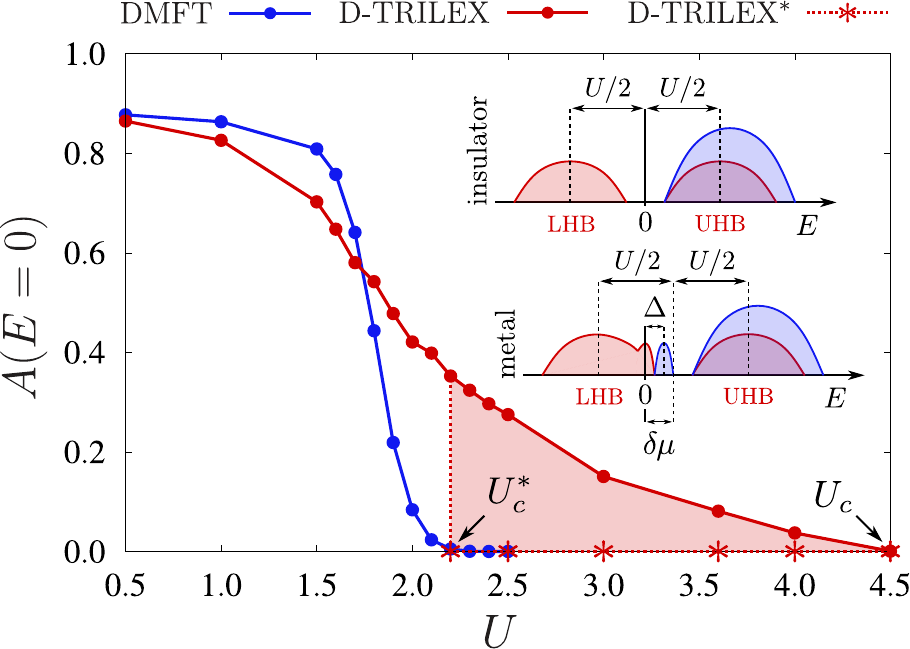}
\caption{Electronic density at Fermi energy ${A(E=0)}$ for the lower orbital (${l=2}$) as a function of the interaction $U$. The result is obtained from DMFT (blue dots), metallic \mbox{D-TRILEX} (red dots), and insulating \mbox{D-TRILEX}$^{*}$ (red asterisks) solutions. The red shaded area highlights the simultaneous existence of the metallic and the Mott insulating solutions.
The inset sketches the difference in the DOS between the insulating (top) and metallic (bottom) \mbox{D-TRILEX} solutions.
In the insulating case, the Fermi energy lies between the LHB and UHB that are split approximately by $U$. 
In the metallic case, the difference in the chemical potential ${\delta\mu=\mu^{*}-\mu}$ brings the upper part of the LHB to the Fermi energy, which results in the formation of the quasi-particle peak at ${E=0}$.
The splitting between the quasi-particle peaks coincides with the value of the crystal field splitting ${\simeq\Delta}$.
\label{fig:Phase}}
\end{figure}

{\it Results.}
To illustrate the effect of non-local correlations on the Mott transition, we compare the DOS predicted by DMFT and \mbox{D-TRILEX} methods.
The result of these calculations is shown in Fig.~\ref{fig:DOS} for three different values of the interaction ${U=1.8}$, ${U=2.0}$, and ${U=2.2}$.
First, let us focus on the quarter-filled calculations presented in the two upper rows of this figure. 
We find that the results of the DMFT and \mbox{D-TRILEX} methods are different already at ${U=1.8}$. 
In both cases, the DOS is metallic.
The lower orbital (${l=2}$, red line) displays a three-peak structure consisting of the quasi-particle peak at Fermi energy ${E=0}$ and two side peaks that correspond to lower and upper Hubbard bands (LHB and UHB).
The upper orbital (${l=1}$, blue line) also exhibits the quasi-particle peak in the DOS that appears close to the Fermi energy at ${E \simeq \Delta}$.
However, the three-peak structure predicted by DMFT possesses a high degree of electron-hole symmetry.
Instead, the DOS of obtained for the same orbital (${l=1}$) using the \mbox{D-TRILEX} approach resembles the DOS of a hole-doped Mott insulator with the quasi-particle peak being shifted closer to the LHB~\cite{RevModPhys.68.13}.
The quasi-particle peaks in the DOS of DMFT vanish simultaneously between ${U=1.8}$ and ${U=2.0}$, which signals the tendency towards a Mott insulating state in a multi-orbital system at finite temperature.
A further increase of the interaction decreases the electronic density at Fermi energy ${A(E=0)}$.
The latter reaches zero at ${U^{\ast}_{c}\simeq2.2}$ (blue line in Fig.~\ref{fig:Phase}), and the DMFT solution enters the Mott insulating phase.
On the contrary, the \mbox{D-TRILEX} solution remains metallic for the discussed values of the interaction (middle row in Fig.~\ref{fig:DOS}).
Thus, even at ${U^{\ast}_{c}}$ it reveals pronounced quasi-particle peaks in the DOS for both orbitals.
Fig.~\ref{fig:Phase} shows that ${A(E=0)}$ in the metallic \mbox{D-TRILEX} solution also decreases upon increasing the interaction.
However, this solution turns into a Mott insulator only at a very strong critical interaction ${U_{c}\simeq4.5}$, which is larger than twice the bandwidth. 
This result seems surprising, since in the single-orbital case the non-local correlations lead to a more insulating electronic behavior~\cite{PhysRevLett.101.186403}, as correctly captured by the \mbox{D-TRILEX} method~\cite{PhysRevB.100.205115}.

\begin{figure}[t!]
\centering
\includegraphics[width=\linewidth]{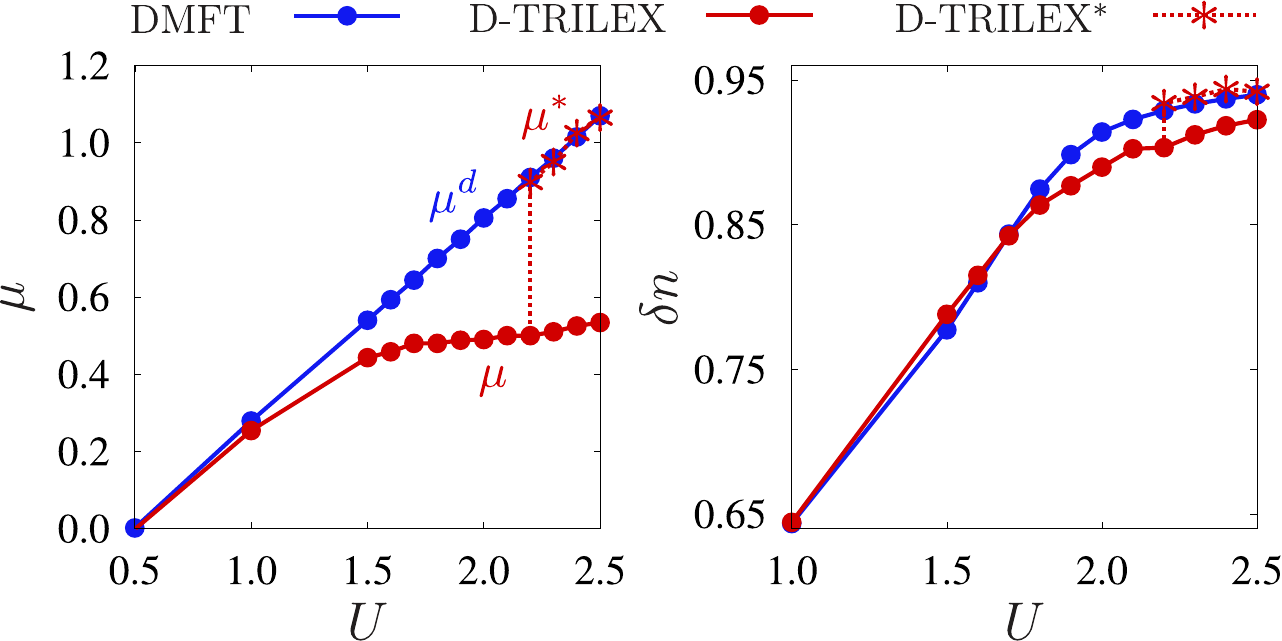}
\caption{Chemical potential (left panel) and orbital polarization (right panel) for the quarter-filled DMFT (blue dots), metallic \mbox{D-TRILEX} (red dots), and insulating \mbox{D-TRILEX} (red asterisks) solutions. The result is obtained for different values of the interaction $U$. Chemical potentials for the insulating \mbox{D-TRILEX} (${\mu^{*}}$) and DMFT (${\mu^{d}}$) solutions nearly coincide. For ${1.0\lesssim U < 2.2}$ no quarter-filled \mbox{D-TRILEX} solution exists near $\mu^{d}$. The chemical potential $\mu$ for the metallic \mbox{D-TRILEX} solution strongly deviates from $\mu^{d}$ at ${U\geq1.5}$. 
\label{fig:MU_dn}}
\end{figure}

\begin{figure}[t!]
\centering
\includegraphics[width=1.\linewidth]{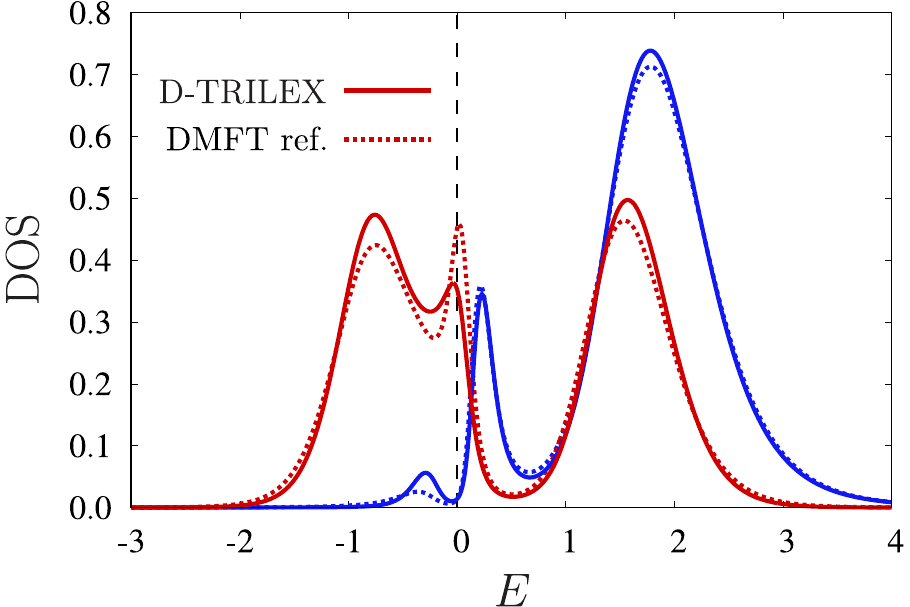}
\caption{DOS for the metallic \mbox{D-TRILEX} solution (solid lines) and its DMFT reference system (dashed lines) obtained for the same value of the chemical potential $\mu$ at the critical interaction ${U^{\ast}_{c}}$. The reference system is a doped Mott insulator with ${\langle n \rangle = 0.96}$. 
\label{fig:DOS_U2.2}}
\end{figure}

To explain the observed effect, we note that quarter-filling in DMFT and \mbox{D-TRILEX} corresponds to different values of the chemical potential.
The left panel of Fig.~\ref{fig:MU_dn} shows that at ${U\geq1.5}$ the chemical potential ${\mu}$ of \mbox{D-TRILEX} (red dots) significantly deviates from $\mu^{d}$ of DMFT (blue dots), and this difference increases with increasing the interaction.
We point out that \mbox{D-TRILEX} calculations are based on the DMFT solution of the local impurity problem that plays a role of the reference system~\cite{PhysRevB.100.205115, PhysRevB.103.245123}.
We find that the quarter-filled metallic \mbox{D-TRILEX} solution originates from the metallic reference system that has smaller average density.
Fig.~\ref{fig:DOS_U2.2} shows that due to ${\langle n \rangle < 1}$ the reference system (dashed lines) remains metallic even at $U^{*}_{c}$.
At the same time, the DOS predicted by \mbox{D-TRILEX} (solid lines) is not dramatically different from the one of the reference system.
This fact suggests that for a given value of the chemical potential the effect of non-local collective electronic fluctuations in the metallic regime consists in moving the spectral weight from above to below the Fermi energy, which brings the filling of the system to ${\langle n \rangle=1}$.

To confirm this statement, we perform \mbox{D-TRILEX} calculations for the chemical potential $\mu^{d}$ of the quarter-filled DMFT solution.
The corresponding result is shown in the bottom row of Fig.~\ref{fig:DOS} and is referred to as the \mbox{D-TRILEX}$^{*}$ calculation in order not to confuse it with the metallic solution.
We observe that the obtained DOS is again practically identical to the one of DMFT (bottom {\it vs.} top row in Fig.~\ref{fig:DOS}).
However, the \mbox{D-TRILEX}$^{\ast}$ calculations performed in the regime ${1.0\lesssim U < 2.2}$, where DMFT solution is metallic, correspond to ${\langle n \rangle>1}$.
Moreover, no quarter-filled \mbox{D-TRILEX}$^{\ast}$ solution is found near ${\mu^{d}}$ in this regime of interactions.
This fact supports our previous finding that in the metallic regime non-local correlations increase the average density of the considered system.

This physical picture changes when the DMFT solution becomes Mott insulating.
We find that the corresponding \mbox{D-TRILEX}$^{*}$ solution undergoes the Mott transition at the same critical interaction ${U^{\ast}_{c}}$ as in DMFT (bottom right panel of Fig.~\ref{fig:DOS}).
Moreover, at ${U\geq{}U^{\ast}_{c}}$ the average density for the \mbox{D-TRILEX}$^{*}$ solution becomes ${\langle n \rangle = 1}$ for ${\mu^{*}\simeq\mu^{d}}$ (bottom left panel of Fig.~\ref{fig:DOS}).
The right panel of Fig.~\ref{fig:MU_dn} shows that the insulating DMFT and \mbox{D-TRILEX}$^{\ast}$ solutions are almost fully polarized and have approximately the same value of $\delta n$, which results in electron-hole symmetric DOS for the lower orbital (top and bottom left panels of Fig.~\ref{fig:MU_dn}).
Consequently, the upper orbital becomes nearly unoccupied and thus cannot strongly interact with the lower one.
Therefore, no transfer of the spectral weight between the orbitals by means of the non-local fluctuations occurs in the insulating regime.
Remarkably, the metallic \mbox{D-TRILEX} solution has a lower $\delta{}n$ compared to DMFT.

At ${U\geq{}U^{\ast}_{c}}$ the \mbox{D-TRILEX}$^{*}$ solution remains quarter-filled and Mott insulating, which is confirmed by the zero electronic density at Fermi energy (red asterisks in Fig.~\ref{fig:Phase}).
Therefore, both, the DMFT and the \mbox{D-TRILEX} methods predict the Mott transition for the considered system at the same value of the critical interaction $U^{\ast}_{c}$. 
However, including non-local collective electronic fluctuations beyond DMFT allows one to additionally capture the metallic solution that coexists with the Mott insulating one up to the second critical interaction $U_{c}$. 
For ${U>U_{c}}$ any value of the chemical potential inside the Mott gap gives the same average density, and the two solutions corresponding to $\mu$ and $\mu^{*}$ can be considered equivalent.
A more detailed discussion of the hysteresis curve appearing in Fig.~\ref{fig:MU_dn} can be found in Supplemental Material (SM)~\cite{SM}.

Coexisting solutions with the same average density but different values of the chemical potential have also been found in the DMFT solution of the Hubbard-Kanamori model for small doping around half-filling~\cite{PhysRevLett.118.167003, PhysRevB.102.205127, Sherman_2020, chatzieleftheriou:tel-03391043, 2022arXiv220302451C}, and for different parameters using a strong-coupling expansion~\cite{Sherman_2020}.
Since the quarter-filled model considered in our work displays a strong orbital polarization, it can be expected that taking into account the Hund's rule coupling $J$, which is present in the Kanamori parametrization of the electronic interaction~\cite{Kanamori63, doi:10.1146/annurev-conmatphys-020911-125045}, should not qualitatively change the observed results.
To confirm this point, we perform calculations for the case of $J=U/6$ and find that the meta-stability discussed above survives also in this case, as shown in SM~\cite{SM}.

{\it Conclusions.}
We investigated the effect of non-local collective electronic fluctuations on the Mott transition in a two-orbital quarter-filled model with density-density interaction by comparing the results of the \mbox{D-TRILEX} and DMFT methods.
At the considered temperature, the DMFT solution of the problem remains metallic below the critical interaction ${U^{\ast}_{c}=2.2}$, and at this value of the interaction undergoes the Mott transition.
We find that the inclusion of non-local correlations by means of the \mbox{D-TRILEX} approach stabilizes the metallic phase up to the very large critical interaction ${U_{c}=4.5}$.
The \mbox{D-TRILEX} method also captures the appearance of Mott insulating phase at ${U^{\ast}_{c}}$ as a second meta-stable solution.
This leads to a remarkably broad coexistence region between the metallic and the Mott insulating phases that exist at the same filling, but with different values of the chemical potential between the $U^{\ast}_{c}$ and the $U_{c}$ critical interactions.
Our results show, that for a simple two-orbital model, DMFT cannot correctly interpolate between the moderately- and strongly-interacting regimes, in analogy with the single-orbital case.
This fact brings further evidence that non-local correlations may lead to non-trivial effects due to the presence of additional channels for collective electronic fluctuations also in multi-orbital systems.

\begin{acknowledgments}
M.V., V.H., and A.I.L. acknowledge the support by the Cluster of Excellence ``Advanced Imaging of Matter'' of the Deutsche Forschungsgemeinschaft (DFG) - EXC 2056 - Project No.~ID390715994, and by North-German Supercomputing Alliance (HLRN) under the Project No.~hhp00042.
J.K. and K.H. have been supported by the Austrian Science Fund (FWF) through projects P32044 and I5868 (``Quast'').
V.H. and A.I.L. further acknowledge the support by the DFG through FOR 5249-449872909 (Project P8) and the European Research Council via Synergy Grant 854843-FASTCORR.
The work of E.A.S. was supported by the European Union’s Horizon 2020 Research and Innovation programme under the Marie Sk\l{}odowska Curie grant agreement No.~839551 - \mbox{2DMAGICS}.
E.A.S. is also thankful to the CPHT computer support team.
\end{acknowledgments}

\bibliography{main}

\end{document}


\title{
Supplemental Material\\[0.5cm]
Extended regime of meta-stable metallic and insulating phases in a two-orbital electronic system
}

\author{M. Vandelli}
\affiliation{I. Institute of Theoretical Physics, University of Hamburg, Jungiusstrasse 9, 20355 Hamburg, Germany}
\affiliation{The Hamburg Centre for Ultrafast Imaging, Luruper Chaussee 149, 22761 Hamburg, Germany}
\affiliation{Max Planck Institute for the Structure and Dynamics of Matter,
Center for Free Electron Laser Science, 22761 Hamburg, Germany}

\author{J. Kaufmann}
\affiliation{Institute of Solid State Physics, TU Wien, 1040 Vienna, Austria}

\author{V. Harkov}
\affiliation{I. Institute of Theoretical Physics, University of Hamburg, Jungiusstrasse 9, 20355 Hamburg, Germany}
\affiliation{European X-Ray Free-Electron Laser Facility, Holzkoppel 4, 22869 Schenefeld, Germany}

\author{A. I. Lichtenstein}
\affiliation{I. Institute of Theoretical Physics, University of Hamburg, Jungiusstrasse 9, 20355 Hamburg, Germany}
\affiliation{European X-Ray Free-Electron Laser Facility, Holzkoppel 4, 22869 Schenefeld, Germany}
\affiliation{The Hamburg Centre for Ultrafast Imaging, Luruper Chaussee 149, 22761 Hamburg, Germany}

\author{K. Held}
\affiliation{Institute of Solid State Physics, TU Wien, 1040 Vienna, Austria}

\author{E. A. Stepanov}
\affiliation{CPHT, CNRS, Ecole Polytechnique, Institut Polytechnique de Paris, F-91128 Palaiseau, France}

\maketitle

\onecolumngrid

\begin{figure}[h!]
\includegraphics[width=0.92\linewidth]{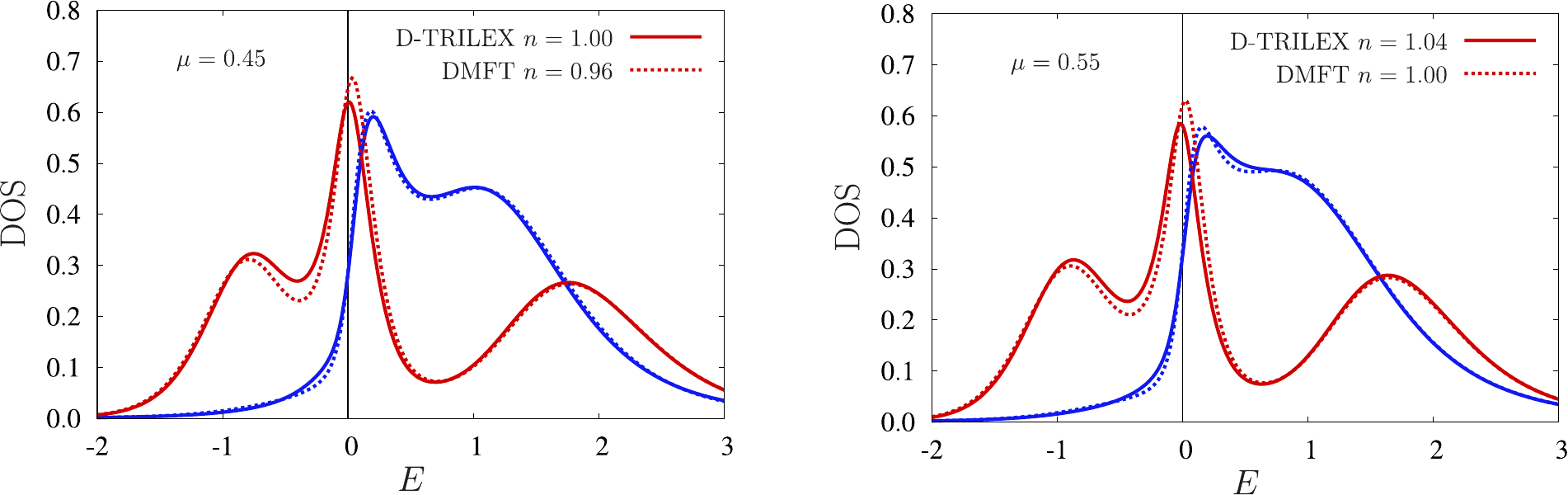}
\caption{\label{fig:U2.4_SM} 
DOS for the upper (${l=1}$, blue line) and lower (${l=2}$, red line) orbitals obtained for ${U=2.4}$ and ${J=U/6}$ using \mbox{D-TRILEX} (solid lines) and DMFT (dashed lines) methods. Left panel corresponds to the chemical potential ${\mu=0.45}$ at which the \mbox{D-TRILEX} solution is quarter-filled (${\langle n \rangle = 1.00}$). Right panel corresponds to the chemical potential ${\mu=0.55}$ of the quarter-filled DMFT solution.}
\end{figure}

\twocolumngrid

{\bf Charge compressibility and meta-stability.}
The meta-stability of the two solutions is signalled by the appearance of two different values of the chemical potential $\mu$ and $\mu^{*}$ leading to the same average density ${\langle n \rangle}$, as demonstrated in the left panel of Fig.~3 in the main text.
As shown there, the presence of two meta-stable solutions manifest itself with the appearance of an hysteresis loop. The branch of the hysteresis chosen by the system depends on whether the corresponding reference system is metallic or Mott insulating. As a matter of fact, if we follow the ${\mu(U)}$ curve that gives ${\langle n \rangle = 1}$ in the weak coupling regime (red dots), we obtain the metallic solution until it continuously turns into an insulating phase at ${U_{c}}$. 
Above this threshold, any value of the chemical potential inside the Mott gap gives the same average density, and the two solutions corresponding to $\mu$ and $\mu^{*}$ can be considered equivalent from there on. 
On the other hand, if we start from the chemical potential $\mu^{*}$ that corresponds to the insulating phase and decrease the interaction following the condition ${\langle n \rangle = 1}$, we obtain the insulating solution (red asterisks). 
The latter exists until the critical interaction ${U^{\ast}_{c}}$ below which no solution for ${\mu^{*}\simeq\mu^{d}}$ is available at quarter-filling.
This behavior means that the function ${\langle n \rangle(\mu)}$ is not monotonic and exhibits a region of negative charge compressibility ${\kappa =  \frac{1}{\langle n \rangle^2}\frac{{\rm d} \langle n \rangle}{{\rm d} \mu}}$. 
According to our calculations, the metallic and insulating \mbox{D-TRILEX} solutions are both characterized by ${\kappa > 0}$, hence they are thermodynamically meta-stable. Since the density is the same for both phases, they have to be separated by a region of chemical potentials associated with a negative charge compressibility.

From the perspective of applications, this region of negative charge compressibility between the two meta-stable solutions could forbid the spontaneous switch between the two solutions. In this case, the application of a static electric field should be sufficient to drive the transition between the metallic and Mott insulating phases, as this perturbation would effectively change the chemical potential from $\mu$ to $\mu^{*}$ or {\it vice versa}. A similar switching between two meta-stable phases as a function of an electric field was investigated in Ref.~\cite{PhysRevLett.114.226403}, where a similar regime of parameters with coexisting metallic and Mott insulating meta-stable solutions was reported. One can speculate that the observed presence of two meta-stable phases could be exploited in the realization of Mott-based electronic switches or transistors. Indeed, the experimental realization of Mott field effect transistors (MottFET) was shown to be practically viable~\cite{doi:10.1063/1.121999, doi:10.1063/1.3651612, Zhong2015}. 
The simultaneous presence of metallic and Mott insulating states could be detected experimentally, for example by measuring dielectric properties of the system~\cite{Pustogow2021}.\\

\begin{figure*}[t!]
\includegraphics[width=0.92\linewidth]{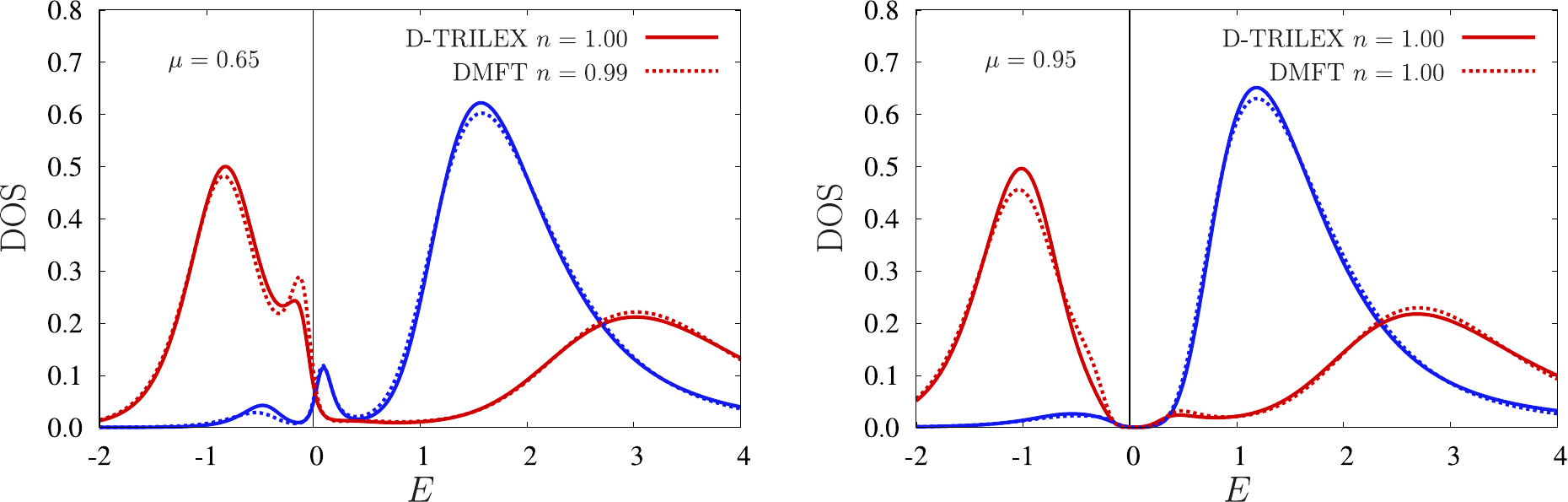}
\caption{\label{fig:U4.2_SM} 
DOS for the upper (${l=1}$, blue line) and lower (${l=2}$, red line) orbitals obtained for ${U=4.2}$ and ${J=U/6}$ using \mbox{D-TRILEX} (solid lines) and DMFT (dashed lines) methods. Left panel corresponds to the chemical potential ${\mu=0.65}$ at which the \mbox{D-TRILEX} solution is metallic and quarter-filled (${\langle n \rangle = 1.00}$). Right panel corresponds to the chemical potential ${\mu=0.95}$ at which both, \mbox{D-TRILEX} and DMFT solutions are Mott insulating and quarter-filled.}
\end{figure*}

{\bf Calculations for a non-zero Hund's coupling.}
The quarter-filled (${\langle n \rangle = 1}$ electrons per lattice site) two-orbital model considered in the main text has a relatively big value of the crystal-field splitting.
This leads to a large orbital polarization that appears already in the metallic regime before the system undergoes the Mott transition.
The latter means that the single electron at each lattice site mostly populates the lower orbital (${l=2}$), and the upper orbital (${l=1}$) stays nearly unoccupied.
In this case, it can be expected that including the Hund's rule coupling $J$ in the electronic interaction should not qualitatively change the physical effect observed in the absence of $J$.
To confirm this point, we perform calculations for the local density of states (DOS) for two different interaction strength ${U=2.4}$ (Fig.~\ref{fig:U2.4_SM}) and ${U=4.2}$ (Fig.~\ref{fig:U4.2_SM}) for a non-zero value of the Hund's coupling ${J=U/6}$.

Due to the effect of $J$, the system is still metallic at ${U=2.4}$ in contrast to the ${J=0}$ case, and we find a single value of the chemical potential $\mu$ for which the \mbox{D-TRILEX} solution has an average density ${\langle n \rangle = 1}$. As in the case of $J=0$ discussed in the main text, the quarter filling in DMFT and \mbox{D-TRILEX} methods corresponds to different values of the chemical potential.
At ${\mu=0.45}$ (left panel in Fig.~\ref{fig:U2.4_SM}), when the \mbox{D-TRILEX} solution is quarter-filled, the average density in DMFT is ${\langle n \rangle = 0.96}$ electrons per lattice site. 
The chemical potential ${\mu=0.55}$ (right panel in Fig.~\ref{fig:U2.4_SM}) corresponds to the quarter-filled DMFT solution, while the average density in \mbox{D-TRILEX} is ${\langle n \rangle = 1.04}$.
We find that both methods predict a rather similar DOS for each of the two values of the chemical potential.
However, the average density of \mbox{D-TRILEX} is always larger than in DMFT. 
This observation is in agreement with the results reported in the main text for the case of ${J=0}$ and illustrates that in the metallic regime the role of the non-local fluctuations in the considered system consists in redistributing the spectral weight between the orbitals.

At ${U=4.2}$ the quarter-filled DMFT solution that corresponds to ${\mu=0.95}$ lies in the Mott insulating regime (right panel in Fig.~\ref{fig:U4.2_SM}).
For this value of the chemical potential the \mbox{D-TRILEX} solution is also quarter-filled and Mott insulating, because no spectral weight redistribution induced by the non-local fluctuations occurs in the insulating regime.
Finally, at smaller value of the chemical potential ${\mu=0.65}$ (left panel in Fig.~\ref{fig:U4.2_SM}) the \mbox{D-TRILEX} reveals the second quarter-filled solution, which is metallic.
The DOS predicted by DMFT for this value of the chemical potential is again similar to the one of \mbox{D-TRILEX}, but the DMFT solution does not reside at quarter filling and corresponds to ${\langle n \rangle = 0.99}$. 
Therefore, we find that when the DMFT solution becomes Mott insulating \mbox{D-TRILEX} reveals two different quarter-filled solutions even for the case of a non-zero value of the Hund's coupling.

\bibliography{main}